\documentclass[journal,twocolumn]{IEEEtran}
\usepackage{cite}
\usepackage{amsmath,amssymb,amsfonts}
\usepackage{algorithmic}
\usepackage{graphicx}
\usepackage{textcomp}
\usepackage{lipsum,mathtools,cuted}
\usepackage{xcolor}

\usepackage{float}
\usepackage{epsfig}
\usepackage{epstopdf}
\usepackage{subfigure,color}
\usepackage{dcolumn}
\usepackage{lineno}
\usepackage{accents}
\usepackage{multirow}
\usepackage[usestackEOL]{stackengine}
\usepackage{tabularx}
\usepackage{makecell}
\usepackage{dcolumn}
\usepackage{bm}
\usepackage{booktabs}
\UseRawInputEncoding

\def\BibTeX{{\rm B\kern-.05em{\sc i\kern-.025em b}\kern-.08em
  T\kern-.1667em\lower.7ex\hbox{E}\kern-.125emX}}
  
\def\e{\begin{equation}}
\def\f{\end{equation}}
\def\_#1{{\bf #1}}

\def\.{\cdot}

\def\=#1{\overline{\overline #1}}

\def\@#1{_{\rm #1}}

\begin{document}
\title{Molding of Reflection and Scattering from Uniform Walls Using Space-Periodic Metasurfaces}

\author{S.~Kosulnikov, F.S.~Cuesta, X.~Wang, and S.A.~Tretyakov,~\IEEEmembership{Fellow, IEEE}

\thanks{(Corresponding author: X.~Wang) This work was supported in part by the European Commission through the Horizon 2020 (H2020) Artificial Intelligence Aided D-band Network for 5G Long Term Evolution (ARIADNE) project under grant 871464 and by the Academy of Finland under grant 345178.}
\thanks{S.~Kosulnikov, F.~S.~Cuesta, and S.~A.~Tretyakov are with the Department of Electronics and Nanoengineering, Aalto University, FI-00076 Aalto, Finland. X.~Wang is with the Institute of Nanotechnology, Karlsruhe Institute of Technology, Karlsruhe,
Germany.
(e-mail: sergei.2.kosulnikov@aalto.fi;
francisco.cuestasoto@aalto.fi;
xuchen.wang@kit.edu;
sergei.tretyakov@aalto.fi).}}

\maketitle

\begin{abstract}
Active development is taking place in reconfigurable and static metasurfaces that control and optimize reflections. 
However, existing designs typically only optimize reflections from the metasurface panels, neglecting interference with reflections originating from supporting walls and nearby objects in realistic scenarios. 
Moreover, when the area illuminated by the transmitting antenna is larger than the metasurface panel, the total scattering pattern deviates significantly from the metasurface panel's reflection pattern. 
In this study, we investigate how engineering the metasurface properties can modify the total scattering pattern, enabling the modification and optimization of reflections from significantly larger illuminated areas than the metasurface panel.
To accomplish this, a general design approach is developed to create periodical metasurfaces with controlled reflection phase and amplitude for arbitrary Floquet channels. 
By combining these reflections with those from the surrounding walls, the total scattering can be manipulated to produce desired scattering properties. 
The study demonstrates how appropriately designed metasurfaces can modify reflections from surrounding walls, enhancing the functionalities of metasurfaces. 
These findings are intended to facilitate advancements in engineering and optimizing wave propagation channels, particularly for millimeter-wave communications.

\end{abstract}
\begin{IEEEkeywords}
Metasurfaces,
Electromagnetic diffraction,
Harmonic analysis, 
Millimeter wave propagation, 
Surface impedance
\end{IEEEkeywords}


\section{Introduction}

Over the past few decades, wireless technologies have significantly advanced, showcasing their potential for improving quality of life.
The constant demand for increasing bandwidth and transfer rates has pushed wireless links technologies towards higher-frequency bands.
However, this trend has posed a challenge, as newer wireless devices must deal with the inherent propagation losses at these higher frequencies. 
One possible solution is the use of highly-directive antennas, which provides gain in the desired direction and compensates for the additional path losses.
Highly directive antennas have already been used in particular engineering areas for a long time, but their use has generally been limited to fixed wireless links, e.g., for satellite communications, or for radio links between  base stations \cite{Rahmat-Samii2015,Xu2020Review,Raut2021}.
Nevertheless, new technological solutions need to be considered when the receiver is a mobile object, as highly-directive antennas in weakly scattering environments provide limited coverage areas.
Therefore, new high-frequency technologies consider wireless channels that are optimized in real time \cite{Jensen2004,Alexiou2004,Arapoglou2011}.
Unfortunately, there are possible cases when the direct line of sight (LOS) connection between the user and the base station is blocked due to some physical obstacle. 
At the same time, very high antenna directivity and high operational frequencies in the millimeter-wave range do not allow us to use the reflected signal paths, as we do nowadays in microwave technologies.

One of the prospective solutions is changing the environment and creating alternative channels that avoid the obstacles. 
The fixed or reconfigurable intelligent surface (RIS) approach is targeted to realizations of this functionality using anomalous reflection \cite{DiRenzo2020}. 
There are different methods to design metasurfaces \cite{Glybovski2016}, in particular, for anomalous reflection (e.g., \cite{Chang_RA,pozar1997design, epstein2016synthesis,asadchy2016perfect, asadchy2017eliminating, Radi_Metagrating_2018, Popov_Controlling_2018, Diaz_Power_2019, Do-Hoon_planar_2021, budhu2021perfectly}). 
In the design of such metasurfaces, it is typically assumed that the structures are infinitely large and illuminated by plane waves. 
However, in many application scenarios such as those found in 6G or beyond wireless communications,  to reduce the size and costs of the metasurface, the metasurface area may be smaller than the illuminated spot on the supporting wall.  In this situation a considerable amount of power is reflected by the supporting wall, ultimately leading to unexpected scattering effects that can undermine the desired performance of the metasurface.

We have recently developed an efficient method for design of anomalous reflectors and experimentally confirmed effective operation for D-band frequencies \cite{Wang2020, kosulnikov2023discrete}.
The method is quite general, and it can be used for the design of Floquet metasurfaces with arbitrary power distributions and arbitrarily selected phases among all the propagating Floquet modes of a periodic metasurface \cite{kosulnikov2023discrete}.
In this work, we will focus on metasurfaces for engineering beam splitting with both power and phase control. 
As an application example, we will use these metasurfaces to modify reflections also from uniform walls surrounding a metasurface panel, which usually form a beam reflected into the specular direction. We show how a properly designed metasurface panel can modify scattering from a significantly wider area than the metasurface itself. 
Here, our focus is on fixed (static) metasurfaces, however, the results can be applied in the future also for reconfigurable reflectors.
This study is most relevant for high-frequency (millimeter-wave and THz) systems, where signals are usually carried by narrow beams, and it is important to be able to engineer scattering patterns from illuminated objects, in order to ensure required coverage. 

This paper is organized as follows: in Section II, we provide a brief description of an effective method to design versatile reflective metasurfaces with advanced functionalities and analyze the efficiency of the implemented infinite models. 
In Section III, we first validate the performance by comparison of  simulated scattering from finite metasurface panels with the theoretical model; second, we extend the analysis to demonstrate how the considered metasurfaces can modify specular reflection from the supporting walls.
Sections IV concludes the paper.

\section{Design of reflectors with advanced functionalities} \label{sec:design}

In paper \cite{kosulnikov2023discrete}, we developed a general analytical approach based on the mode-matching method for calculating and optimizing all the scattering harmonics of an arbitrary periodic metasurface composed of discrete impedance patches positioned on a grounded substrate, given a certain incidence at a specific frequency and incident angle. 
Here, we will not repeat the derivations and directly use the method for the design of metasurfaces with multiple engineered Floquet channels. 

\begin{figure}[!h]
\centerline{\includegraphics[width=0.8\linewidth]{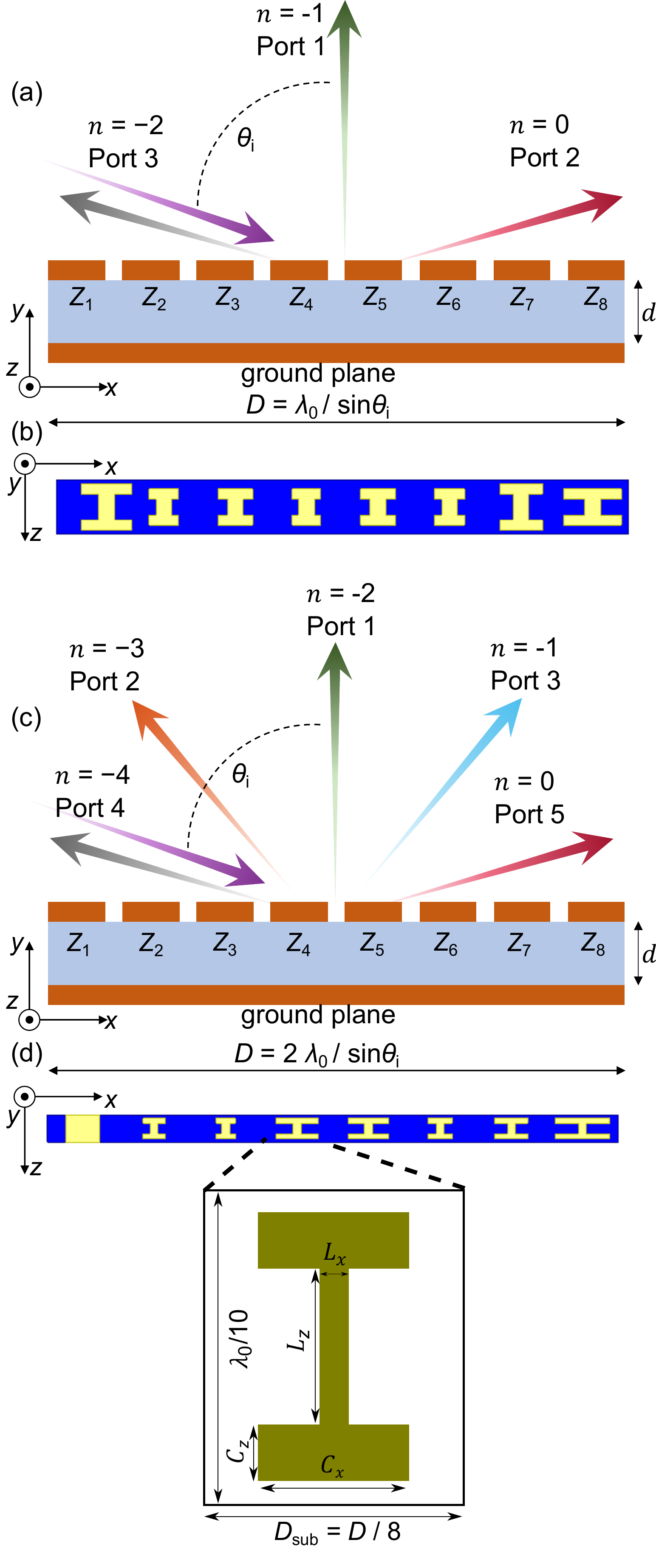}}
\caption{Unit cells of the metasurfaces under study: (a) -- schematic side view of a 3 Floquet port metasurface designed as an anomalous reflector, a 3-channel splitter, and as a specular reflector with an engineered reflection phase; (b) -- top view of the implemented pattern for the 3-channel splitter; (c) -- schematic side view of the 5-channel splitter; (d) -- top view of the implemented 5-channel splitter. }
\label{fig:ChanScatt}
\end{figure}

At  the first design step, the metasurface is modeled as a periodic array of supercells formed by  impedance patches ($Z_1$, $Z_2$, $\cdots$, $Z_K$) on a grounded dielectric substrate, as shown on Fig.~\ref{fig:ChanScatt}(a). 
We use the method presented in  \cite{kosulnikov2023discrete} to obtain the scattering amplitude and phase into each Floquet channel. Each channel corresponds to one of the propagating Floquet harmonics, and the angle of reflection for the $n$-th harmonic $\theta_n$ is determined by the expression
\begin{equation}
  \sin\theta_n=\sin\theta_{\rm i}+\frac{2 \pi n}{k D},
  \label{Floquet_period}
\end{equation} 
where $\theta\@{i}$ is the incidence angle, $k$ is the wavenumber of the incident wave, and $D$ is the period (the supercell size) \cite[Chapter~7]{Ishimaru1991}. 
Considering the scattering harmonics from $n=-N$ to $n=+N$, the amplitudes of the incident and reflected harmonics can be represented by $2N+1$ dimensional arrays, $\mathbf{E}_{\rm i}$ and $\mathbf{E}_{\rm r}$, respectively. 
These arrays are related by the reflection matrix $\mathbf{R}$, which is a $2N+1$ square matrix. 
The reflection matrix is determined by the substrate parameters and the discrete sheet-impedance values $Z_1, Z_2, \cdots Z_K$ of the impedance patches.

The design of metasurfaces is an inverse problem, where the goal is to find a proper set of discrete impedance values for a given incidence and desired reflection harmonics. 
After the sheet impedances of the patches are known, they can be realized as properly shaped metal patches. Since there is no analytical solution for this inverse problem, we use  a mathematical optimization \cite{kosulnikov2023discrete}. 
For excitations by a single TE-polarized plane wave, the incident-field vector $\mathbf{E}_{\rm i}$ can be written as an array with all zeros except for the central element which is equal to unity,
\begin{equation}
\mathbf{E}_{\rm i}=[0, \cdots, 0, 1, 0, \cdots, 0]^T.
\end{equation}
Knowing the incident field $\mathbf{E}_{\rm i}$, the amplitudes of all scattered modes can be found as $\mathbf{E}_{\rm r}= \mathbf{R}\cdot \mathbf{E}_{\rm i}$,
\begin{equation}
\mathbf{E}_{\rm r}=[R_{-N} \cdots, R_{-1}, R_{0}, R_{1}, \cdots, R_{N}]^T,
\end{equation}
where $R_n$  refers to the complex amplitude of $n$-th scattering harmonic.
The optimization target is to find a proper set of sheet impedance values $Z_k$ to ensure that the calculated $\mathbf{E}_{\rm r}$ satisfies the design purposes. 
A schematic side view of the diffracted harmonics scattered from three-channel metasurfaces is shown in Fig.~\ref{fig:ChanScatt}(a).

For example, we can require that in the design of three-channel beam splitters the incident wave from port 3 is equally split to port 2 ($n=0$) and port 1 ($n=-1$), and the reflection wave from port 1 has zero reflection phase. 
In this case, we define the following optimization goals:
\begin{equation}
  |R_0|=0.707, \quad |R_{-1}|=\frac{\sqrt{2}}{2}\sqrt{\frac{\cos\theta_{\rm i}}{\cos\theta_{-1}}}, \quad \arg(R_0)=0 . \label{eq: objective}
\end{equation}
The first two objectives ensure that the power is equally distributed between ports 1 and 2. The third one ensures zero reflection phase at port 2. The reflection angle for illumination from port 3 to port 1 (corresponding to the Floquet harmonic $n=-1$) should correspond to  the direction of interest $\theta_{-1}=\theta_{\rm r}$, which is ensured by tuning the size of the supercell $D$ according to Eq.~\eqref{Floquet_period}.
We optimize the discrete impedance values $Z_1, Z_2, \cdots, Z_K$ (purely reactive) until the calculated reflection parameters meet the objectives. 
To find the optimal values of the sheet impedances, we use the mathematical optimization tool available in the MATLAB package.
At each optimization step, MATLAB assumes an array of $Z_1, Z_2, \cdots, Z_K$ and calculates the reflected fields. 
Then, the cost is evaluated by summing up the differences between the desired and actual reflection parameters. 
Using the MultiStart and $fmincon$ algorithms, MATLAB searches for the minimum value of the cost function in the multidimensional parameter space.

In what follows, using the above optimization method, we find the required sheet impedances $Z_1, Z_2, \cdots, Z_K$ for different metasurfaces with advanced functionalities, controlling the amplitude and phases of multichannel metasurfaces. To illustrate versatile possibilities of multi-beam reflection control, we design four types of metasurfaces, namely: a perfect anomalous reflector, a three-channel splitter with zero specular reflection phase, an artificial magnetic conductor (AMC) working as a specular reflector with zero reflection phase, and a five-channel splitter with reflected waves towards two anomalous directions. 
Furthermore, we present actual implementations of the designed metasurfaces as arrays of ``dog-bone'' metal patches. We do not implement the anomalous reflector since an equivalent structure was already implemented and experimentally validated in Ref.~\cite{kosulnikov2023discrete}. The AMC is also straightforward to implement, as it allows a direct analytical solution and has been investigated in detail in many papers (see e.g. \cite{sievenpiper1999high}). We use all of the designed surfaces to discuss possibilities to shape reflections also from the illuminated part of the supporting wall.

In all considered examples, the structures operate at 144.75~GHz (the corresponding free-space wavelength $\lambda_0 \approx 2.072$~mm), and the oblique incidence is from $\theta_{\rm i} = 70^{\circ}$ for all four structures, as presented in Fig.~\ref{fig:ChanScatt}(a,c). 
The periodic metasurface of the period $D$ is discretized to $K = 8$ subcells, therefore, the subcell size along the $x$-axis is $D_{\rm{sub}} = D/8$.
The metasurfaces are realized as three-layer structures made of a penetrable infinitely thin impedance sheet over a Quartz substrate with $\varepsilon_{\rm r} = 4.2$ and the loss tangent 0.005, see an illustration on Fig.~\ref{fig:ChanScatt}(a).
The substrate thickness is $d = 209.5~\mu$m, and the ground plane is made of copper. In the following of this section, we present the designs of the four metasurfaces with the assumption of infinite size. The scattering properties of finite-sized metasurface will be investigated in Section~III.

\subsection{Anomalous reflector}

Our first reference structure is an extreme case when all the incident power is reflected towards an anomalous reflection direction -- an anomalous reflector. 
We set $\theta_{\rm i} = 70^{\circ}$ and optimize the structure to maximize the amplitude of the $n=-1$ reflected harmonic. 
The optimization objective is the value corresponding to perfect conversion of the incident power to this mode: $|R_{-1}|=\sqrt{{\cos\theta_{\rm i}}/{\cos\theta_{\rm r}}}$ \cite{epstein2016synthesis,asadchy2016perfect}. 
The optimization gives the reactive sheet impedance values $Z = [-132, -278, -187, -1215, -1099, -1008, -989, 50]j$~Ohm.
First we test the optimization results by ANSYS HFSS, simulating an infinite periodic array   with periodic boundary conditions. We assume a finite thickness and realistic losses of the dielectric substrate, a finite conductivity of the background metal, and the patch arrays are modelled as penetrable impedance sheets (i.s.). 
The corresponding scattering matrix $S_{\rm i.s.}$ reads
\begin{equation}
|S_{\rm i.s.}| = 
  \begin{bmatrix}
    {0.01 } & {0.01 } & {\mathbf{0.99} }\\ 
    {0.01 } & {0.98 } & {0.02 } \\
    {0.99 } & {0.02 } & {0.00 }
  \end{bmatrix}.
\end{equation}

From the scattering matrix of the simulated structure we see a confirmation of the desired functionality: the structure works as an effective anomalous reflector. 
Anomalous reflection from this infinite structure is almost perfect, with the efficiency $\eta_{\rm eff} = |S_{13}|^2 = 99 \%$. 
The imperfections are caused mainly by losses in the dielectric substrate and in the copper ground plane. Worth noting that such structure can be directly implemented as it was already demonstrated and validated in Ref.~\cite{kosulnikov2023discrete}.

\subsection{Three-channel splitter}

In the optimization process, we keep the same periodicity $D=\lambda_0/\sin \theta_{\rm i}$, and set $\theta_{\rm i} = 70^{\circ}$. 
As a target, we request the optimization objective as defined in Eq.~(\ref{eq: objective}). 
As a result of the optimization we get the corresponding sheet impedance vector 
$Z = [-611, -262, -911, -806, -948, -771, -951, -209]j$~Ohm.
The scattering matrix resulting from the EM simulation of an infinite structure formed by an array of these impedance sheets at a lossy dielectric substrate $S_{\rm i.s.}$ reads
\begin{equation}
  S_{\rm i.s.} = 
   \resizebox{.38\textwidth}{!}{ $ \begin{bmatrix}
    {0.51 \angle -24.6^{\circ}} & {0.49 \angle 89.9^{\circ}} & \mathbf{0.69 \angle -114^{\circ}} \\ 
    {0.49 \angle -90.1^{\circ}} & {0.48 \angle 23.5^{\circ}} & \mathbf{0.69 \angle -0.1^{\circ}} \\
    {0.69 \angle 65.9^{\circ}} & {0.69 \angle -0.1^{\circ}} & {0 \angle -133^{\circ}}
  \end{bmatrix}$}.
\end{equation}
From this scattering matrix we see that the incident power is split between the 1st and 2nd ports very precisely, and the reflection phases towards the specular direction ($S_{23}$ and $S_{32}$) are close to zero.

The next target for this splitter structure design is to find appropriate physical elements for implementation of these impedance sheets. 
We use a procedure similar to that presented in Ref.~\cite{kosulnikov2023discrete} to find ``dog-bone'' meta-atom dimensions granting the desired sheet impedance values. 
This type of meta-atoms allows to achieve reliable angular stability. 
The discretization step of the structure along the $z$-axis is $\lambda_0 / 10$. 

We implement the optimized sheet impedance one by one and combine them together. This direct implementation and consideration of lossess in both finite conductivity copper metal patches and the dielectric substrate grant  the scattering matrix $S_{\rm imp}^{\rm dir}$ in the EM simulation as follows:
\begin{equation}
  S_{\rm imp}^{\rm dir} = 
   \resizebox{.37\textwidth}{!}{ $\begin{bmatrix}
    {0.47 \angle -24.4^{\circ}} & {0.46 \angle 90.8^{\circ}} & \mathbf{0.69 \angle -112^{\circ}} \\ 
    {0.46 \angle -89.2^{\circ}} & {0.49 \angle 25^{\circ}} & \mathbf{0.59 \angle 7.58^{\circ}} \\
    {0.69 \angle 68.1^{\circ}} & {0.59 \angle 7.58^{\circ}} & {0.03 \angle 163^{\circ}} 
  \end{bmatrix}$}.
\end{equation}
From these results we see that the power splitting is less perfect, and also the specular reflection phase increases to $7.58^{\circ}$. 
Due to strong near-field interactions, the structure is sensitive to the implemented sheet impedance values, and it should be accurately fine-tuned for the best performance. 
Therefore, we implement additional numerical tuning of the meta-atom parameters in order to equate the split powers and minimize the specular reflection phase. 
The optimized design parameters are presented in Table~\ref{tab:MainTable} (column ``3 chan.'') in accordance with the structural parameters shown in the inset of Fig.~\ref{fig:ChanScatt}. 
Worth noting that we shifted the first dog-bone element inside the subcell by $0.2$ of the subcell period to improve the desired functionality.
The overall top view of the implemented supercell is presented in Fig.~\ref{fig:ChanScatt}(b).

\begin{table}
\centering
\caption{Design parameters of the implemented splitters. 
}
\begin{tabular}{||c | c | c ||}
\hline
& 3 chan. & 5 chan. \\ [0.5ex] 
\hline\hline
$C_{x1}$, $\mu$m & 223 & 420 \\
\hline
$C_{x2}$, $\mu$m & 164 & 257 \\
\hline
$C_{x3}$, $\mu$m & 116 & 188 \\
\hline
$C_{x4}$, $\mu$m & 134 & 316 \\
\hline
$C_{x5}$, $\mu$m & 110 & 330 \\
\hline
$C_{x6}$, $\mu$m & 136 & 147 \\
\hline
$C_{x7}$, $\mu$m & 114 & 170 \\
\hline
$C_{x8}$, $\mu$m & 194 & 0 \\
\hline
$C_{z1}$, $\mu$m & 40 & 40 \\
\hline
$C_{z2}$, $\mu$m & 40 & 40 \\
\hline
$C_{z3}$, $\mu$m & 40 & 40 \\
\hline
$C_{z4}$, $\mu$m & 40 & 40 \\
\hline
$C_{z5}$, $\mu$m & 40 & 40 \\
\hline
$C_{z6}$, $\mu$m & 40 & 40 \\
\hline
$C_{z7}$, $\mu$m & 40 & 40 \\
\hline
$C_{z8}$, $\mu$m & 40 & 0 \\
\hline
$L_{x1}$, $\mu$m & 60 & 60 \\
\hline
$L_{x2}$, $\mu$m & 60 & 60 \\
\hline
$L_{x3}$, $\mu$m & 60 & 60 \\
\hline
$L_{x4}$, $\mu$m & 60 & 60 \\
\hline
$L_{x5}$, $\mu$m & 60 & 60 \\
\hline
$L_{x6}$, $\mu$m & 60 & 60 \\
\hline
$L_{x7}$, $\mu$m & 60 & 60 \\
\hline
$L_{x8}$, $\mu$m & 60 & 263 \\
\hline
$L_{z1}$, $\mu$m & 60 & 80 \\
\hline
$L_{z2}$, $\mu$m & 100 & 80 \\
\hline
$L_{z3}$, $\mu$m & 60 & 80 \\
\hline
$L_{z4}$, $\mu$m & 60 & 80 \\
\hline
$L_{z5}$, $\mu$m & 60 & 80 \\
\hline
$L_{z6}$, $\mu$m & 60 & 80 \\
\hline
$L_{z7}$, $\mu$m & 60 & 80 \\
\hline
$L_{z8}$, $\mu$m & 100 & $\lambda_0 /10$ \\
\hline
\end{tabular}
\label{tab:MainTable}
\end{table}

Considering both losses in the substrate dielectric and the finite conductivity of metal elements, the scattering matrix of the implemented optimized infinite structure $S_{\rm imp}^{\rm opt}$ reads
\begin{equation}
 S_{\rm imp}^{\rm opt} = 
  \resizebox{.37\textwidth}{!}{ $ \begin{bmatrix} 
    {0.55 \angle -30.3^{\circ}} & {0.45 \angle 92^{\circ}} & \mathbf{0.64 \angle -105^{\circ}} \\ 
    {0.46 \angle -88^{\circ}} & {0.47 \angle -8.48^{\circ}} & \mathbf{0.62 \angle 0.53^{\circ}} \\
    {0.64 \angle 75^{\circ}} & {0.62 \angle 0.53^{\circ}} & {0.18 \angle -115^{\circ}}
  \end{bmatrix}$}.
\end{equation}
This result confirms the desired performance. The power is nearly equally split, and the specular reflection phase is negligibly small. 
However, at the same time the retroreflection coefficient $S_{33}$ grows up to $3.2 \%$.

\subsection{Artificial Magnetic Conductor}

The considered design approach can also be used to design uniform metasurfaces, e.g., here, an AMC.
It works as a specular reflector with zero phase delay.
We proceed with the 3-port functionality  [see Fig.~\ref{fig:ChanScatt}(a)].
In this example, we set the incidence angle $\theta_{\rm i} = 70^\circ$ and maximize the amplitude of the $n=0$ scattered harmonic, with the target value $R_0 = 1$.
As expected, the best solution from the optimization code results with almost identical values for all subcells' sheet impedances $Z_{1-8} = -472j$~Ohm. 
The resulting sheet reactance is in accordance to that obtained from the equivalent transmission-line model of a homogeneous grid on a grounded substrate.

The scattering matrix given by the EM simulation of an infinite structure modeled as an impedance sheet at a lossy dielectric $S_{\rm i.s.}$ reads
\begin{equation}
  S_{\rm i.s.} = 
  \resizebox{.38\textwidth}{!}{ $ \begin{bmatrix}
    {0.99 \angle -26.4^{\circ}} & {0.00 \angle -10.7^{\circ}} & {0.00 \angle -14.9^{\circ}} \\ 
    {0.00 \angle 169^{\circ}} & {0.00 \angle 129^{\circ}} & \mathbf{0.97 \angle 0.21^{\circ}} \\
    {0.00 \angle 165^{\circ}} & {0.97 \angle 0.21^{\circ}} & {0.00 \angle 49.7^{\circ}}
  \end{bmatrix}$}.
\end{equation}
This result confirms the desired functionality. 
Slight imperfections in the reflected power are caused by dissipation due to losses in the dielectric and the ground-plane materials. We do not go further with implementation of this structure, since the same implementation procedure as demonstrated here for the splitters (or for anomalous reflectors in Ref.~\cite{kosulnikov2023discrete}) can be easily applied in this example.

\subsection{Splitting into two anomalous directions}

In order to split the energy between two anomalous directions we double the periodicity of the structure, setting it to $D = 2 \lambda_0/\sin{\theta_{\rm i}}$. 
For illumination from $\theta_{\rm i}=70^{\circ}$, there are the following propagating reflected harmonic angles:
$\theta_{0} = 70^{\circ}$,
$\theta_{-1} = 28.024^{\circ}$,
$\theta_{-2} = 0^{\circ}$,
$\theta_{-3} = -28.024^{\circ}$,
$\theta_{-4} = -70^{\circ}$.
In this example, we define the target to  equally split the incident power to the $n=-1$ and $n=-2$ anomalous scattering harmonics. The objective is $|R_{-2}| = \sqrt{\frac{1}{2}\frac{\cos{\theta_{\rm i}}}{\cos{\theta_{\rm -2}}}} $, and $|R_{-1}| =\sqrt{\frac{1}{2}\frac{\cos{\theta_{\rm i}}}{\cos{\theta_{\rm -1}}}} $.

As a result of the optimization of the sheet impedances for the desired 5-channel splitter, we obtain the vector of values 
$Z = [-110, -427, -662, -294, -265, -867, -750, 40]j$~Ohm.
The scattering matrix of an array of impedance sheets, considering the dielectric material losses and the finite conductivity of the ground plane reads $S_{\rm i.s.}$
\begin{equation*}
  |S_{\rm i.s.}| = 
  \begin{bmatrix}
    {0.31 } & {0.47 } & {0.32} & \textbf{0.69 } & {0.29 } \\ 
    {0.47 } & {0.65 } & {0.44} & {0.00 } & {0.39 } \\
    {0.32 } & {0.44 } & {0.32 } & \textbf{0.71 } & {0.30 } \\
    {0.69 } & {0.00 } & {0.71} & {0.00 } & {0.01 } \\
    {0.29 } & {0.39 } & {0.30} & {0.01 } & {0.80 }
  \end{bmatrix}.
\end{equation*}
This result confirms the desired functionality: $S_{14} \approx S_{34}$, whereas the other scattering parameters for illumination from port 4 are negligibly small.

Next we present the implementation procedure, which is similar to that for the 3-channel splitter. 
The geometrical parameters of the implemented dog-bone meta-atoms are presented in Table~\ref{tab:MainTable} in the column ``5 chan.'', in accordance with the structural parameters in the inset of Fig.~\ref{fig:ChanScatt}. 
The overall top view of the implemented structure is given in Fig.~\ref{fig:ChanScatt}(d). 
The scattering matrix of the 5-channel splitter with the implemented meta-atoms (direct implementation, no optimization), considering losses in the dielectric substrate and finite conductivity of the metal pattern and the background $S_{\rm imp}$ reads
\begin{equation}
  |S_{\rm imp.}|= 
  \begin{bmatrix}
    {0.43 } & {0.32 } & {0.29 } & \textbf{0.61 } & {0.38 } \\ 
    {0.32 } & {0.57 } & {0.34 } & {0.18 } & {0.52 } \\
    {0.29 } & {0.34 } & {0.40 } & \textbf{0.61 } & {0.21 } \\
    {0.61} & {0.18 } & {0.61 } & {0.18} & {0.06} \\
    {0.38 } & {0.52 } & {0.21 } & {0.06 } & {0.45 }
  \end{bmatrix}.
\end{equation}
The results confirm the desired functionality.
Further manual optimization did not provide any considerable enhancement of performance.

\section{Scattering patterns of metasurface panels on uniform walls} \label{sec:MathEst}

In this section, we apply the developed metasurfaces' designs to validate their scattering performance with the consideration of reflecting walls.
We validate full-wave EM simulations with previously developed theoretical model of scattering from the finite-sized metasurfaces \cite{kosulnikov2023simple}. 
Finally we analyze how these structures may be utilized for molding scattering towards specular direction based on the theoretical estimations.

\begin{figure}[ht]
\centerline{\includegraphics[width=0.8\linewidth]{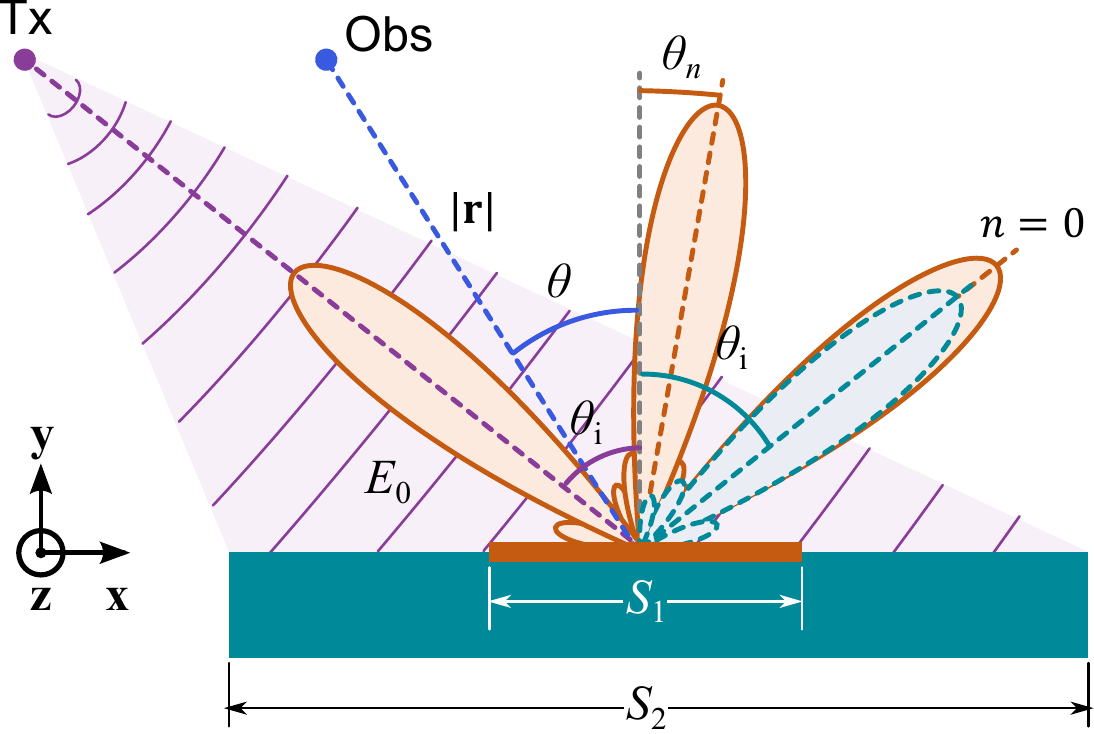}}
\caption{Schematic of a metasurface of the area $S_1$ on top of a uniform wall. The illuminated area equals $S_2$. }
\label{fig:ChanScatt3}
\end{figure}

Let us consider a metasurface with the supercell configuration of Fig.~\ref{fig:ChanScatt}. 
If the metasurface pattern repeats infinitely, the scattering produced by plane-wave illumination is a combination of plane-wave Floquet harmonics with specular and non-specular propagation directions, following  Eq.~\eqref{Floquet_period}. 
In a more realistic scenario, illustrated in Fig.~\ref{fig:ChanScatt3}, the metasurface has a finite area $S_1$ and it is illuminated by a directive beam. 
We assume that in the far zone of the transmitting antenna the wavefront can be approximated by a plane wave. 
The metasurface is placed at a wall with homogeneous properties, and the illuminated surface area is $S_2$. 
For TE-polarized incidence ($z$-polarized in the present notations), the scattered field in the far zone reads \cite{Diaz-Rubio2022}
\begin{equation}
\begin{split}
  {E}_{{\rm sc} z} & = 
  \frac{jk}{4\pi} {\frac{e^{-jk|\mathbf{r}|}}{|\mathbf{r}|}}E_0 \Bigg[ S_2 \Bigg( (1+R\@{wall})\cos\theta\\
  & -(1-R\@{wall})\cos\theta\@{i} \Bigg) {\rm sinc}(ka\@{ef})\\
  & +S_1\sum_n(R_n-R\@{wall}\delta_n)(\cos\theta+\cos\theta_{n}){\rm sinc}(ka_{{\rm ef}, n}) \Bigg].
\end{split}  
  \label{final_reflective}
\end{equation}
Here, $\mathbf{r}$ is the position vector pointing to the observation point, $R\@{wall}$ is the reflection coefficient of the surrounding uniform wall, $R_n$ is the macroscopic reflection coefficient for the $n$-th Floquet harmonic of the metasurface currents, $\theta$ is the angle between the normal and the direction to the observation point, $a\@{ef} = (\sin \theta - \sin \theta\@{i})a_2/2$, $a_{{\rm ef}, n} = (\sin \theta - \sin \theta_{n})a_1/2$, $\delta_n$ is the Kronecker delta ($\delta_0=1$ and $\delta_{n\neq 0}=0$).

Four subsections below discuss performance of all considered metasurface examples realized as finite-sized samples.
First, we validate the scattering performance of the finite-size array, comparing the full-wave simulation model with the theoretical model presented in Eq.~\eqref{final_reflective}, calculating the scattered electric field ${E}_{{\rm sc} z}$ at 1~m distance from the metasurface. 
These simulations are computationally demanding, therefore we use semi-analytical models, modeling patch arrays by sheet impedances and considering relatively small finite-size samples. 
We compare two cases: 1) when the illuminated area covers the metasurface only; 2) when the illumination covers the metasurface and a part of the surrounding reflecting wall. 
We consider the surrounding wall as a PEC boundary as the most extreme case with the reflection coefficient $R_{\rm wall} = -1$, however, the used theoretical model is applicable for any value of the wall reflection coefficient \cite{kosulnikov2023simple}. 
The total sizes of the simulated structures are $S_1 = 8D \times 8D$ and $S_2 = 16D \times 16D$ (with an illuminated wall) or $S_2 = S_1$ (only a metasurface), in accordance with the model in Eq.~\eqref{final_reflective}. 
The metasurfaces are square panels of the area $S_1 = a_1^2$. 
Similarly to the semi-analytical EM simulations provided for the infinitely periodic structures in Sec.~\ref{sec:design}, the finite metasurfaces are modeled in ANSYS HFSS as three-layer structures made of penetrable impedance sheets over a Quartz substrate, and the ground plane made of copper. These results are presented in Fig.~\ref{fig:TheoryAndSimCart}.

Next, we utilize the same theoretical model of Eq.~\eqref{final_reflective} to analyse scattering from these four metasurfaces toward the specular region. To this end, we  fix the illuminated spot area to $S_2 = a_2^2 = 100 \lambda_0 \times 100 \lambda_0$ and calculate the scattered fields at 1~m distance in a narrow angular sector close to the specular direction. 
Similarly to our full-wave simulations, the metasurfaces are located at a PEC wall. 
The structures in this theoretical analysis are of a larger electrical size to be closer to prospective application  scenarios in indoor environment.
The main goal here is to study how the specular reflection from the wall is modified due to reflections from metasurface panels of different sizes in the range $0 \leq a_1 \leq a_2$. These results are presented in Fig.~\ref{fig:MathEst}.

\subsection{Anomalous reflector}

\begin{figure}
\centerline{\includegraphics[width=0.8\linewidth]{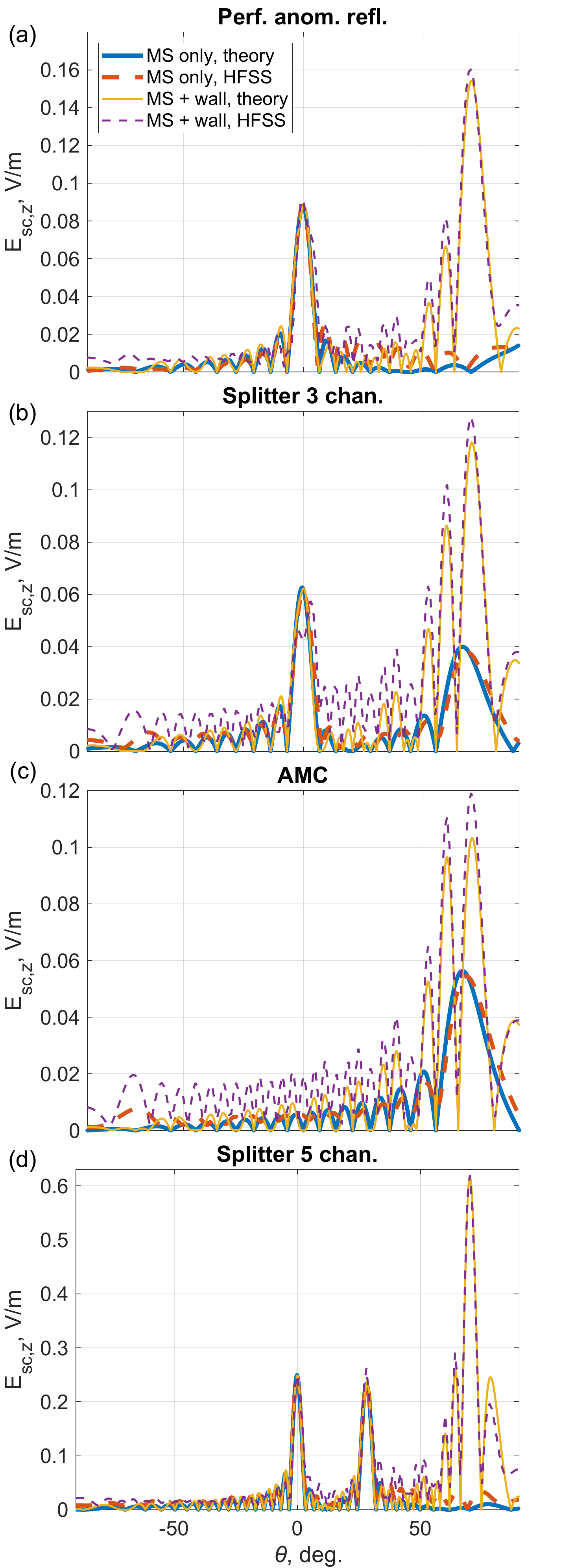}}
\caption{Comparison of the scattered electric field $E_{{\rm sc} z}$ at 1~m distance from the metasurfaces under plane-wave illumination at $\theta_{\rm i} = -70^{\circ}$ calculated using the theoretical model Eq.~(\ref{final_reflective}) and EM simulations: (a) -- perfect anomalous reflector; (b) -- splitter 3-channel; (c) -- perfect specular reflector; (d) -- splitter 5-channel. For every case we compare scattering from the metasurface only and from the metasurface surrounded by a PEC wall. Note that in case (d) the areas $S_1$ and $S_2$ are larger, because the period $D$ is twice as large as compared to cases (a-c).}
\label{fig:TheoryAndSimCart}
\end{figure}

\begin{figure*}[ht]
\centerline{\includegraphics[width=1\textwidth]{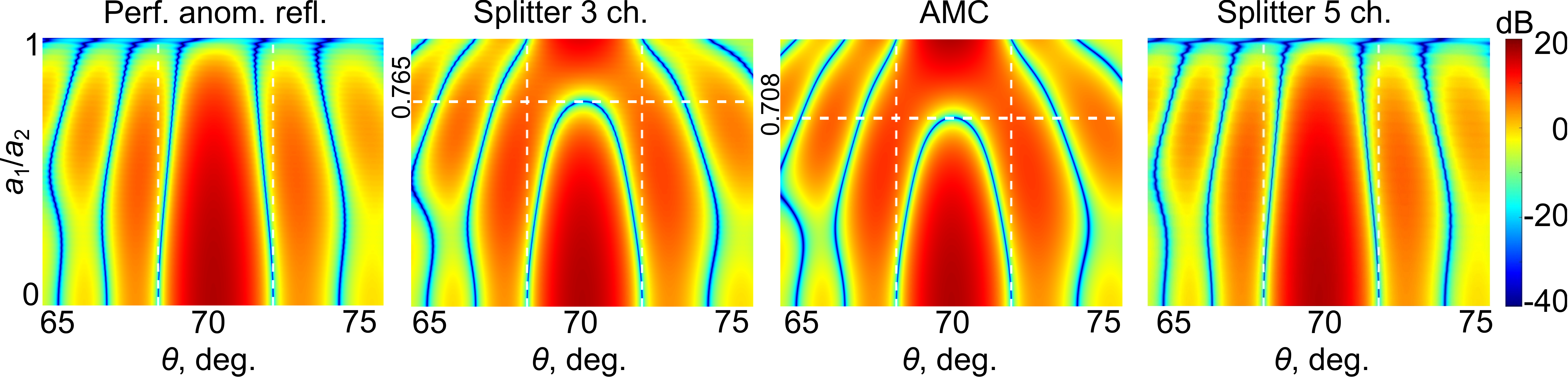}}
\caption{Scattering pattern in a small sector around the specular reflection angle. The structures are described in Sec.~\ref{sec:MathEst}. The scattered field is calculated using the model of Eq.~\eqref{final_reflective}: perfect anomalous reflector; three-channel splitter with $0^{\circ}$ specular reflection phase; perfect specular reflector with $0^{\circ}$ specular reflection phase; and five-channel splitter into two anomalous directions. $S_2 = 100 \lambda_0 \times 100 \lambda_0$, plane-wave illumination at $\theta_{\rm i}=70^\circ$. The field is calculated at the distance of 1~m $\approx 482.6\lambda_0$ from the structures. The vertical white dashed lines show the initial width of the specular lobe, and the Horizontal dashed lines show the ratio region where the specular scattering is suppressed.}
\label{fig:MathEst}
\end{figure*}

The first structure under consideration is an anomalous reflector, which is a three-port structure [see Fig.~\ref{fig:ChanScatt}(a)] where the amplitude of the reflected mode $n=-1$ is maximized. 
The far-field scattering pattern of this metasurface panel is shown in Fig.~\ref{fig:TheoryAndSimCart}(a).
The theoretical result using Eq.~\eqref{final_reflective} and considering only the presence of the metasurface (without illumination of the surrounding PEC wall) is marked as ``MS only, theory''.
The corresponding result from the EM simulation of the same finite metasurface is marked as ``MS only, HFSS''.
In a realistic scenario, the total scattered field at the receiver position is a combination of the power scattered by the metasurface and the power reflected from the wall. 
Therefore, the other two lines marked as ``MS + wall'' show the case when the illuminated area covers also a part of the surrounding wall.  
One can see from the results that the anomalous reflection towards $0^{\circ}$ remains close to the estimated, as the metasurface area remains the same. There is only some modification of side-lobe scattering. However, significant scattering towards the specular direction appears for the case when the surrounding wall is illuminated. 

As seen in Eq.~\eqref{final_reflective}, the amplitudes of the scattered fields produced by the metasurface and the illuminated wall are proportional to their areas. 
Results in Fig.~\ref{fig:MathEst}(a) show how the field intensity in the specular direction $\theta_{0}=70^\circ$ is affected by the metasurface size for varying illuminated area of the wall. 
Vertical white dashed lines show the initial width of the specular lobe, which is similar for all four considered cases, when the metasurface is absent.
An increment of the metasurface area leads to stronger destructive interference in the specular reflection, resulting in complete suppression when the metasurface size $a_1$ is equal to the illuminated area of the wall. 

However, in this example, modifications of the specular beam   pattern are small, because only weak side-lobe radiation from the metasurface interferences with the wall reflection.
%
In the next example, we investigate how a phase-controlled beam splitter can be used to partly suppress specular reflection and create significant diffuse scattering from uniform support walls.

\subsection{Three-channel splitter}

The designed three-port splitter [see Fig.~\ref{fig:ChanScatt}(a)] divides the scattered power equally between the specularly reflected mode $n=0$ and the anomalously reflected mode $n=-1$. 
Additionally, the metasurface is designed to produce specular reflection with zero phase delay. 
The metasurface performs as intended, as seen in Fig.~\ref{fig:TheoryAndSimCart}(b). 
Specular reflection from the surrounding wall is in this case out of phase with the reflection from the metasurface, as revealed in the results from Fig.~\ref{fig:MathEst}(b).
The results are very different from the anomalous reflector case, due to destructive interference between specular reflections from the metasurface and the surrounding wall. 
The specularly reflected lobe degrades faster, compared to the anomalous reflector, and its width shrinks. 
The most noticeable effect is found at $a_1/a_2 \approx 0.765$ (marked with horizontal white line) where the metasurface and the illuminated part of the wall reflect the same power towards the specular direction, resulting in a pattern null at $\theta_{\rm i}$. 
At the same time, amplitudes of the side lobes around the specular direction increase, producing diffuse scattering effects.
When the specular reflection from the metasurface prevails over the wall reflection, we see how the two neighbouring maxima merge into a new wide lobe towards specular direction.

\subsection{Artificial magnetic Conductor}

Even stronger effects of PEC wall-reflection control are seen if the metasurface is designed as an artifical  magnetic conductor. 
To calculate the corresponding patterns,  we set $R_{ n \neq 0} = 0$ and $R_{ n = 0} = 1$ in Eq.~\eqref{final_reflective}, which gives the scattering pattern presented in Fig.~\ref{fig:TheoryAndSimCart}(c). 
The results of the equivalent analysis as for the anomalous reflector with varying $a_1/a_2$ are presented in Fig.~\ref{fig:MathEst}(c). 
We see that, as expected, the effect of specular scattering lobe vanishing and side-lobe enhancement occurs now faster as compared to the three-channel power splitter. 
Indeed, in this case all the incident power at the metasurface area is used to affect the reflection from the illuminated part of the wall. 
The null in the specular reflection occurs at $a_1/a_2\approx 0.708$ (marked with the horizontal white line). 
Thus, the considered metasurface controls reflection from the area that is in total about twice as large as the metasurface itself ($S_2 \approx 2S_1$).

\subsection{Splitting into two anomalous directions}

The last structure analysed in this work is a five-channel metasurface, shown in Fig.~\ref{fig:ChanScatt}(c). The main goal of considering this example is to demonstrate the versatility of the design method. Here, we maximize scattering into two anomalous directions for oblique-incidence illumination (the $n=-1$ and $n=-2$ harmonics). 
Reflection into the $n=-1$ harmonic produces non-specular scattering in the direction $\theta_{-1}={\rm asin} \left(0.5\sin \theta_{\rm i}\right)$. According to Eq.~\eqref{Floquet_period}, this corresponds to radiating towards $\theta_{-1}\approx 28^\circ$ in the case that $\theta\@{i}= 70^\circ$. The other anomalous reflection beam $n=-1$ is in the normal direction. In this case, reflections from the metasurface into the specular directions are small, similarly to the three-channel splitter, but one of the two reflected beams can be used to control reflections from the supporting wall if the wall is not  flat surface. For example, when the metasurface is mounted close to a wall corner.  

The results with only the metasurface in Fig.~\ref{fig:TheoryAndSimCart}(d) demonstrate the intended properties of the structure.
As is seen from Fig.~\ref{fig:MathEst}(d), there is very little effect of the metasurface on reflections into the specular direction from this flat wall. 
This is similar to the anomalous reflector, with minor changes in the corresponding values of $a_1 \rightarrow a_2$ due to somewhat different metasurface scattering into the side-lobes. 
This is clear, because the design of both these metasurfaces minimizes any possible reflections towards the specular direction.

In summary, the results from Fig.~\ref{fig:MathEst} bring an alternative path for metasurface applications, where the desired scattering is produced not only by the metasurface itself but also due to engineered interference with the surrounding environment. 
In the considered example of the use of destructive interference with specular reflections from the uniform support wall, we see that it is possible to reduce specular reflection, if that is not desirable, and enhance scattering into other directions, enhancing diffuse scattering in the propagation channel in the millimeter-wave band. 
Importantly, we see that the overall area of controllable reflection is significantly larger than the area of the metasurface. 

\section{Conclusions}

In this paper we discussed possibilities to mold reflected fields from illuminated spots on uniform walls using advanced anomalous reflectors. 
The study is relevant to millimeter-wave telecommunications where directive antennas are used and the propagation environment does not offer rich diffuse scattering. 
In the considered examples we introduced metasurfaces that create several controllable reflected beams and enhance diffuse scattering. 
Engineering the multichannel behavior of the considered structures allows us to effectively shape reflections into the specular direction and into the side-lobes. 
In particular, it was shown that metasurfaces can effectively modify specular reflections from surrounding walls, redirecting a certain part of the specularly scattered power into secondary lobes. 

The paper describes the complete design process, from the definition of the desired functionality to finding the appropriate shape and dimensions of meta-atoms and finally to estimations of the far-field scattering pattern of the metasurface placed at a uniform wall. 
In the considered examples we assumed perfect conductivity of the supporting wall, which required zero-phase specular reflection from metasurface in order to realize destructive interference. 
However, the demonstrated possibility to arbitrarily engineer the reflection amplitude and phase for all propagating Floquet harmonics allows us to achieve the same effects for walls with arbitrary reflection coefficients. 
Results of an experimental test of a millimeter-wave anomalous reflector designed using the developed method can be found in \cite{kosulnikov2023discrete}.

We believe that these results are useful for optimization of propagation channels in prospective telecommunication systems utilizing high-frequency bands. 
Reconfigurable versions of similar multi-channel reflectors can enhance application scenarios of engineered surfaces. 

\bibliographystyle{IEEEtran}
\bibliography{IEEEabrv,references}

\end{document}